# Spark-ignited kernel dynamics in fine ethanol sprays and their relations with minimum ignition energy

Qiang Li and Huangwei Zhang[*]

*Department of Mechanical Engineering, National University of Singapore, 9 Engineering Drive 1, Singapore 117576, Republic of Singapore*

**Abstract**

Spark ignition of ethanol droplet/vapor/air mixture is studied with a Eulerian-Eulerian method and detailed chemical mechanism. The flame kernel-droplet interaction is quantified with an evaporation completion front (ECF). Two categories of spray flames can hence be defined based on the relative location between the ECF and flame front, i.e., homogeneous and heterogeneous spray flames. An element-based equivalence ratio (ER) at the flame front (flame ER for short) is introduced to measure the gas composition in evaporating sprays. For overall fuel-lean mixtures, quasi-stationary spherical flame (QSSF) occurs due to lean flame ER and the composition at the QSSF front is homogeneous. For overall fuel-rich two-phase mixtures, re-ignition, after the spark-ignited kernel fails, is observed when the droplet diameter is 15 $\mu$m for fuel sprays with both fuel-lean and fuel-rich background gas. This is due to rich flame ER and/or strong evaporative heat loss. Meanwhile, the kernel is born in a heterogeneous mixture and transition into homogeneous state is found. For both overall lean and rich two-phase mixtures, fuel droplets affect the ignitability and flame trajectories. Moreover, ignition energy affects the flame ER and front distance at the early stage of kernel development. Lastly, the minimum ignition energies (MIE) with different gas and overall ERs are investigated. Three regimes (A, B and C) are identified from the MIE variations with overall ER and the corresponding flame kernel dynamics behind them are summarized. Regime A is characterized by the QSSF phenomenon, whilst regime C embodies the ignition failure and re-ignition transients when the droplet size is relatively large. Furthermore, regime B only appears with a narrow range of overall ER when initial background gas ER is above unity. These regimes are further generalized in parameter space of overall ER versus initial background gas ER.

---

[*] Corresponding author. Email: huangwei.zhang@nus.edu.sg, Tel: +65 6516 2557.

# Nomenclature

| | | | |
|---|---|---|---|
| $A_d$ | Surface area of a single droplet [m²] | $R$ | Universal gas constant [J/mol/K] |
| $AF_s$ | Molecular ratio of air/fuel under stoichiometric condition for gas mixture | $Re_d$ | Droplet Reynolds number |
| $a_i$ | Constants related to fuel saturation pressure | $R_c$ | Evaporation completion front [m] |
| $B_M$ | Mass transfer number | $R_f$ | Flame front [m] |
| $B_T$ | Heat transfer number | $r$ | Radial coordinate [m] |
| $b_i$ | Constants related to the latent heat of vaporization | $r_{ig}$ | Spark radius [m] |
| $C_d$ | Drag coefficient | $r_w$ | Computational domain length [cm] |
| $C_{p,d}$ | Heat capacity at constant pressure for fuel droplet [J/kg/K] | $Sc$ | Sherwood number |
| $C_{p,v}$ | Heat capacity at constant pressure for fuel vapor [J/kg/K] | $S_b$ | Flame propagation speed [m/s] |
| $D_{ab}$ | Vapour mass diffusivity in the gas phase [m²/s] | $Sh$ | Schmidt number |
| $d$ | Droplet diameter [m] | $S_I$ | Ignition source term |
| $E$ | Total energy of gas phase [J/m³] | $S_L$ | Liquid phase source term |
| $E_{ig}$ | Ignition energy [J] | $S_R$ | Chemical reaction source term [kg/m³/s] or [kg/m²/s²] or [W/m³] |
| $F_s$ | Drag force [N] | $S_e$ | Energy exchange source term [W/m³] |
| HRR | Heat release rate [W/m³] | $S_{m,i}$ | Mass exchange source term for $i$-th species [kg/m³/s] |
| $H_g(T_d)$ | Enthalpy of ethanol vapor at the droplet temperature [J/kg] | $S_v$ | Momentum exchange source term [kg/m²/s²] |
| $h$ | Total enthalpy of gas [J/kg] | $T_{cr}$ | Critical temperature of fuel [K] |
| $h_c$ | Convective heat transfer coefficient [W/m²/K] | $T_d$ | Droplet temperature [K] |
| $k_g$ | Thermal conductivity of gas [W/m/K] | $T_r$ | Temperature ratio, defined as $T_d/T_{cr}$ |
| $Le$ | Lewis number of gas phase | $T_s$ | Vapor temperature at the droplet surface [K] |
| $L_v(T_d)$ | Latent heat of vaporization at the droplet temperature [J/kg] | $T_{sat}$ | Saturation temperature [K] |
| $M_d$ | Molecular weight of droplet [kg/mol] | $t$ | Temporal coordinate [s] |
| $M_{ed}$ | Mean molecular weight of the gas mixture excluding the fuel vapor [kg/mol] | $U$ | Conservative variables |
| $M_g$ | Molecular weight of gas mixture [kg/mol] | $u_d$ | Droplet velocity [m/s] |
| $\dot{m}$ | Evaporation rate of a single droplet [kg/s] | $u_g$ | Gas velocity [m/s] |
| $m_d$ | Mass of a single droplet [kg] | $V_i'$ | Diffusion velocity of $i$-th species [m/s] |
| $N_d$ | Number density of droplet [m$^{-3}$] | $V_1, V_2$ | Constants related to the fuel vapor mass diffusivity |
| $N_u$ | Nusselt number | $X_C$ | Local element mole fraction of C |
| $n$ | Total number of species | $X_{ds}$ | Ethanol vapor mole fraction at the droplet surface |
| $Pr$ | Prandtl number | $X_H$ | Local mole fraction of H element |
| $p_g$ | Gas pressure [Pa] | $X_O$ | Local mole fraction of O element |
| $p_f$ | Partial pressure for fuel vapor in the gas mixture [Pa] | $Y_{ds}$ | Ethanol vapor mass fraction at the droplet surface |
| $p_s$ | Fuel vapor pressure at droplet surface [Pa] | $Y_{d\infty}$ | Ethanol vapor mass fraction in the bulk gas |
| $p_{sat}$ | Saturation pressure for fuel vapor [Pa] | $Y_i$ | Mass fraction of $i$-th species in gas mixture |
| $q_{ig}$ | Ignition energy source term [W/m$^3$] | | |

**Greek letters**

| | | | |
|---|---|---|---|
| $\Delta R$ | Difference between evaporation completion front and flame front [m] | $\phi_l$ | Liquid fuel equivalence ratio |
| $\rho_d$ | Density of fuel droplet [kg/m$^3$] | $\phi_{ov}$ | Overall equivalence ratio |
| $\rho_g$ | Density of gas mixture [kg/m$^3$] | $\tau_1, \tau_2$ | Viscous stress [Pa] |
| $\rho_s$ | Density of fuel vapor at the droplet surface [kg/m$^3$] | $\tau_{ig}$ | Spark duration time [m] |
| $\Phi$ | Viscous dissipation rate of energy [W/m$^3$] | $\tau_r$ | Droplet momentum relaxation time [s] |
| $\phi_e$ | Effective equivalence ratio | $\mu_g$ | Dynamic viscosity of gas [Pa·s] |
| $\phi_g$ | Equivalence ratio of gas mixture | $\omega_i$ | Chemical reaction rate for $i$-th species [kg/m$^3$/s] |

**Subscripts**

| | |
|---|---|
| $d$ | Properties related to droplet |
| $g$ | Properties related to gas |
| $f$ | Properties related to flame front |
| $s$ | Properties related to droplet surface |
| $0$ | Properties related to initial state |

# 1 Introduction

Flame−droplet interaction in spray combustion is a critical topic in both fundamental combustion theory and applied combustion study [1–3]. Due to existence of the dispersed fuel droplets, spray flames exhibit peculiar structures and/or dynamic behaviors different from gaseous flames. For instance, Mizutani et al. [4] and Akamatsu et al. [5] observed a two-layer structure, consisting of a nonluminous flame front and luminous flamelets behind the flame front. The latter is caused by post-flame droplet burning. The two-layer spray flame structure is also observed by Fan et al. [6] when acetone droplets are added to the methane/air mixture. According to the direct numerical simulations by Neophytou et al. [7], a triple flame is observed, which results from the high-concentration fuel vapor from droplet evaporation around the spark. Moreover, flame wrinkling induced by droplet evaporation near the flame front is reported by Ozel et al. [8]. Three distinctive modes of spray flame propagation are identified by de Oliveira and Mastorakos [9] with schlieren and OH* chemiluminescence methods, including droplet, inter-droplet and gaseous-like modes. Flame wrinkling and droplet penetration to the post-flame zone occur in the first two modes with a small Group combustion number ($< 10^2$). Recently, Li and Zhang [10] find quasi-stationary flame kernel and failure/re-ignition transient in spark ignition of ethanol sprays. They also conclude that these two phenomena are related to heat and mass exchange from droplet evaporation.

Interactions between the flame front and evaporating sprays are directly influenced by their spatial distributions. Generally, two scenarios exist [9,11–16]: (I) fuel droplets only exist in the pre-flame zone; (II) fuel droplets exist in both pre- and post-flame zones. In scenario I, the droplets are fully vaporized before the flame front, whereas in scenario II the droplets cannot complete the evaporation in the pre-flame zone and hence penetrate through the flame front. Their occurrence is influenced by the droplet properties. For instance, scenario I is observed for relatively fine droplets (acetone: 5 $\mu$m [6]; *iso*-octane: 10 $\mu$m [17]), whereas scenario II occurs for coarse droplets (acetone: 65 – 75 $\mu$m [18]). Moreover, both scenarios are found in Jet-A spray flame with varied droplet sizes (16 – 33 $\mu$m) [9,19]. They typically result in different spray flame behaviors. For instance, flame front cellularization only occurs in scenario II according to Thimothée et al. [20]. Meanwhile, flame speed oscillation is correlated with the periodical transition between scenario I and II [15]. Most of the foregoing studies target spray flame propagation. Recently, to capture the intrinsic evolutions of two scenarios in a spray ignition process, Li et al. [14] develop a theoretical model through introducing the concepts of evaporation onset and completion fronts, and hence the interactions between the flame front and the above two can be quantified. The results show that the igniting kernel transits from scenario II to I and change of the distributions of droplets relative to the reaction front leads to different kernel evolutions. Nonetheless, due to the limitations of theoretical analysis, how the two scenarios evolve under various spray conditions is not analyzed therein, which merits further studies with detailed numerical simulation and/or high-resolution experimental measurements.

There have been a large body of investigations on minimum ignition energy (MIE) of spray flames. The experiments by Rao and Lefebvre [21], Ballal and Lefebvre [22–25], and Danis et al. [26] show that the MIE monotonically increases with droplet diameter when the latter is relatively large, i.e., > 30 $\mu$m. For smaller droplets, the MIE increases with decreased diameter [2,27]. Therefore, there exists an optimal diameter corresponding to the smallest MIE, e.g., around 25 $\mu$m for overall fuel-lean *n*-heptane spray flames [26]. Moreover, the mixture composition, parameterized by gas ($\phi_g$) and liquid ($\phi_l$) equivalence ratios, also affects the MIE. For instance, for pure sprays ($\phi_g = 0$), the MIE monotonically decreases with $\phi_l$ for kerosene [21–23], *iso*-octane [24,25] and *n*-heptane [26] sprays with fixed droplet size. However, other researchers have reported the existence of an optimal $\phi_l$ in their studies, corresponding to the smallest MIE [28–30]. Furthermore, for partially pre-vaporized sprays ($\phi_g \neq 0$), the gas phase equivalence ratio, $\phi_g$, also influences the MIE [31,32]. Nonetheless, the intrinsic relations between MIE variation and flame kernel dynamics in fuel sprays have not been correlated in previous studies.

Based on the above consideration, we aim to study spark ignition of ethanol droplet/vapor/air mixtures with detailed chemistry. The influences of gas phase and liquid phase properties will be examined, including droplet diameter, gas and liquid equivalence ratios. The objectives of this work include: (1) flame kernel development in fuel sprays and comparison with gaseous flames; (2) flame kernel-droplet interactions; and (3) correlation between flame kernel development modes and MIE variations. The rest of the paper is structured as below. The mathematical model is presented in Section 2, whilst the physical model and numerical implementations are listed in Section 3. The results will be discussed in Section 4, followed by the key conclusions in Section 5.

## 2  Mathematical model

Spark ignition of two-phase ethanol droplet/vapour/air mixtures is simulated with Eulerian−Eulerian method. An in-house reactive flow solver with detailed chemistry and species transport properties, A-SURF [33,34], is used. The accuracies of A-SURF in simulating both gaseous and spray flames have been validated in previous studies [10,34,35]. The governing equations and models are presented below.

## 2.1 Gas phase

The equations for multi-species reactive flows in one-dimensional spherical coordinate read

$$\frac{\partial U}{\partial t} + \frac{\partial F(U)}{\partial r} + 2\frac{G(U)}{r} = F_v(U) + S_R + S_L + S_I. \tag{1}$$

Here $t$ and $r$ are time and radial coordinate, respectively. The vector $U$ is the conservative variables, whilst $\partial F(U)/\partial r$, $2\,G(U)/r$, and $F_v(U)$ are the convection, geometry, and diffusion terms, respectively. $S_R$ and $S_I$ denote chemistry and numerical spark terms, respectively. Detailed descriptions for $U$, $F(U)$, $G(U)$, $F_v(U)$ and $S_R$ in Eq. (1) can be found in previous studies [33,34]. The effects of fuel droplets on the gas phase are considered through the source/sink terms $S_L$, i.e.,

$$S_L = \left[S_{m,1}, S_{m2}, \ldots, S_{m,i}, \ldots, S_{m,n}, S_v, S_e\right]^T. \tag{2}$$

In Eq. (2), $S_L$ includes the species species ($S_{m,i}$), momentum ($S_v$) and energy ($S_e$) exchanges. They are

$$S_{m,i} = \begin{cases} N_d \dot{m} & for \quad ethanol \\ 0 & for\ other\ species \end{cases}, \tag{3}$$

$$S_v = -N_d m_d \frac{u_g - u_d}{\tau_r}, \tag{4}$$

$$S_e = -N_d h_c A_d (T_g - T_d) + N_d \dot{m} H_g(T_d). \tag{5}$$

Here $H_g(T_d)$ is the enthalpy of ethanol vapor at the droplet temperature, $N_d$ is the number density of the fuel droplet. $\dot{m}$ is the evaporation rate of a single droplet, whilst $m_d$ is the mass of a single droplet. $u_d$ and $T_d$ are the droplet velocity and temperature, respectively. $\tau_r$ is droplet momentum relaxation time. $h_c$ is the convective heat transfer coefficient. $A_d$ is the surface area of a single droplet as $A_d = \pi d^2$, where $d$ is the droplet diameter.

## 2.2 Liquid phase

The Eulerian approach is applied to describe the dispersed spray droplets, and equations of droplet size, velocity, temperature, and number density are solved. Note that mono-sectional method is employed and therefore all the droplets of the relevant sprays are represented by one section. Spherical droplets are considered, and droplet temperature is assumed to be uniform due to their small Biot number [36,37]. Moreover, droplet breakup and/or deformation are not considered due to fine droplets and weak aerodynamic fragmentation effects.

The evolution of droplet diameter, $d$, is governed by

$$\frac{\partial d}{\partial t} + u_d \frac{\partial d}{\partial r} = -\frac{2\dot{m}}{\pi \rho_d d^2}, \tag{6}$$

where $\rho_d$ is droplet material density, taking 783.54 kg/m³ for ethanol. The droplet evaporation rate $\dot{m}$ is modeled as [38]

$$\dot{m} = \pi d \rho_s D_{ab} Sh \ln(1 + B_M), \tag{7}$$

where $\rho_s = p_s M_d / RT_s$ is the vapor density at the droplet surface. $p_s$ and $T_s$ are the vapor pressure and temperature at the droplet surface, respectively. $M_d$ is the molecular weight of ethanol vapor. $T_s$ is estimated through $T_s = (2T_d + T_g)/3$ [38], in which $T_g$ is the gas temperature.

The Spalding mass transfer number $B_M$ is

$$B_M = \frac{Y_{ds} - Y_{d\infty}}{1 - Y_{ds}}, \tag{8}$$

where $Y_{d\infty}$ is the ethanol vapor mass fraction in the bulk gas. $Y_{ds}$ is the ethanol vapor mass fraction at the droplet surface, i.e.,

$$Y_{ds} = \frac{M_d X_{ds}}{M_d X_{ds} + M_{ed}(1 - X_{ds})}. \tag{9}$$

Here $M_{ed}$ is the mean molecular weight of the gas mixture excluding the ethanol vapor. $X_{ds}$ is the ethanol vapor mole fraction at the droplet surface

$$X_{ds} = \frac{p_{sat}(T_d)}{p_g}, \tag{10}$$

where $p_g$ is the pressure for gas mixture. $p_{sat}$ is the saturated vapor pressure and is estimated as a function of droplet temperature $T_d$ [39]

$$p_{sat}(T_d) = exp\left(a_1 + \frac{a_2}{T_d} + a_3 \ln T_d + a_4 T_d^{a_5}\right). \tag{11}$$

The constants $a_1$, $a_2$, $a_3$, $a_4$, and $a_5$ are 59.769, -6595.0, -5.0474, 6.3× $10^{-7}$ and 2.0, respectively [39]. Similarly, the vapor pressure at the droplet surface $p_s$ is estimated with the droplet surface temperature $T_s$, i.e.,

$$p_s(T_s) = exp\left(a_1 + \frac{a_2}{T_s} + a_3 \ln T_s + a_4 T_s^{a_5}\right). \tag{12}$$

Moreover, in Eq. (8), the vapor mass diffusivity in the gas mixture, $D_{ab}$, is modeled as [40]

$$D_{ab} = 3.6059 \times 10^{-3} \frac{(1.8 \times T_s)^{1.75}}{p_g} \sqrt{\frac{1}{M_d \times 10^3} + \frac{1}{M_g \times 10^3}} / \left(V_1^{\frac{1}{3}} + V_2^{\frac{1}{3}}\right)^2. \tag{13}$$

The constants $V_1$ and $V_2$ are 141.78 and 20.1 respectively for ethanol and air mixture [41].

The Sherwood number $Sh$ in Eq. (7) is estimated from [38]

$$Sh = 2.0 + \frac{1}{f(B_M)}\left[(1 + Re_d Sc)^{1/3} \max(1, Re_d)^{0.077} - 1\right]. \tag{14}$$

In the above, $f(B) = \ln(1 + B)/B \cdot (1 + B)^{0.7}$ is used to model the change of film thickness due to Stefan flow effects [38]. The Schmidt number $Sc$ of the gas phase is

$$Sc = \frac{\mu_g}{\rho_g D_{ab}}. \tag{15}$$

In Eq. (14), the droplet Reynolds number $Re_d$ is defined as

$$Re_d \equiv \frac{\rho_g d |u_g - u_d|}{\mu_g}, \tag{16}$$

where $\mu_g$ is the dynamic viscosity of the gas mixture.

The equation of droplet velocity takes the following form

$$\frac{\partial u_d}{\partial t} + u_d \frac{\partial u_d}{\partial r} = \frac{F_s}{m_d}. \tag{17}$$

Note that only drag force $F_s$ is considered in our work and it is modelled from $F_s = m_d/\tau_r \cdot (u_g - u_d)$, where $m_d = \rho_d \pi d^3/6$. $\tau_r$ is determined from [42]

$$\tau_r = \frac{\rho_d d^2}{18 \mu_g} \frac{24}{C_d Re_d}. \tag{18}$$

The drag coefficient $C_d$ is modelled as [42]

$$C_d = \begin{cases} \frac{24}{Re_d}\left(1 + \frac{1}{6}Re_d^{2/3}\right), & if\ Re_d \leq 1000 \\ 0.44, & if\ Re_d > 1000 \end{cases}. \tag{19}$$

Evolution of the droplet temperature is governed by

$$m_d C_{P,d}\left(\frac{\partial T_d}{\partial t} + u_d \frac{\partial T_d}{\partial r}\right) = h_c A_d (T_g - T_d) - \dot{m} L_v(T_d), \tag{20}$$

where $C_{P,d}$ is the specific heat capacity of the liquid ethanol. $L_v(T_d)$ is the latent heat of vaporization at the droplet temperature and estimated from [43]

$$L_v(T_d) = b_1 \cdot (1 - T_r)^{[(b_2 \cdot T_r + b_3) \cdot T_r + b_4] \cdot T_r + b_5}, \tag{21}$$

where the constants $b_1$, $b_2$, $b_3$, $b_4$, and $b_5$ are 958345.09 J/kg, -0.4134, 0.75362, 0, and 0 respectively for liquid ethanol [43]. $T_r$ is defined as $T_r = T_d/T_{cr}$, where $T_{cr}$ is the critical temperature and is 516.25 K for ethanol.

In Eq. (20), the convective heat transfer coefficient $h_c$ is

$$h_c = \frac{Nu k_g}{d}, \tag{22}$$

where $k_g$ is the thermal conductivity of the gas mixture. $Nu$ is the Nusselt number, estimated with Rans and Marshall model [44]

$$Nu = 2.0 + \frac{1}{f(B_T)}\left[(1 + Re_d Pr)^{1/3} \max(1, Re_d)^{0.077} - 1\right], \tag{23}$$

where $Pr$ is the Prandtl number of the gas phase and assumed to be unity in this study. $B_T = (1 + B_M)^\varphi - 1$ is the Spalding heat transfer number, in which $\varphi = (C_{p,v}/C_{P,d})/Le$. $Le$ is the Lewis number of the gas mixture and $C_{p,v}$ is the constant pressure specific heat of ethanol vapor.

The equation of droplet number density $N_d$ reads

$$\frac{\partial N_d}{\partial t} + \frac{\partial (N_d u_d)}{\partial r} + 2\frac{N_d u_d}{r} = 0. \tag{24}$$

## 3 Physical model description and characterization

Spark ignition in fine ethanol sprays is simulated with a one-dimensional configuration. The computational domain length is $0 \leq r \leq r_w$, in which $r_w$ is 20.48 cm. Adaptive mesh refinement is used to accurately capture the reactive front and droplet evaporation completion front with relatively low cost. A refinement level of 8 is used, leading to the finest cell size of 16 µm. Operator splitting approach is applied to sequentially calculate the chemistry and convection/diffusion terms. The time integration, diffusive flux, convective flux, and liquid properties ($T_d$, $u_d$, $d$ and $N_d$) are calculated respectively by the Euler explicit, second-order central differencing, MUSCL-Hancock, and first-order upwind schemes [45]. A detailed mechanism (57 species and 383 reactions) [46] is considered for ethanol combustion, which has also been used in previous studies for ethanol spray flames [47–49].

The initial gas flow velocity is zero, i.e., $u_{g,0} = 0$ m/s. The domain is initially filled with a heterogeneous mixture of ethanol droplets, vapor, and air. The reaction system is characterized by liquid fuel and gaseous vapor Equivalence Ratios (ER), i.e., $\phi_g$ and $\phi_l$. The overall ER can be calculated from $\phi_{ov} = \phi_g + \phi_l$. In our simulations, $\phi_g$ ranges from 0.3 to 1.5, while $\phi_{ov}$ is from 0.8 to 4.0. The initial pressure is $p_{g,0} = 1$ atm. The initial gas temperature, $T_{g,0}$, is assumed to be the saturation temperature $T_{sat}$, with which the partial pressure of pre-vaporized ethanol vapor in a gas mixture with a specific equivalence ratio $\phi_g$ is equal to the corresponding saturation vapor pressure of $p_{sat}(T_{sat})$. This ensures that the evaporation of ethanol droplets before the flame front is inhibited, and therefore the vapor release is fully caused by the flame−droplet interactions. The partial pressure of the ethanol vapor, $p_{0,f}$, before sparking is

$$p_{0,f} = \frac{\phi_g}{\phi_g + AF_s} p_{g,0}, \qquad (25)$$

where $AF_s$ is the air/fuel molar ratio for a stoichiometric gas mixture, which is approximately 14.28 for ethanol/air mixture. For $\phi_g$ studied in this paper (0.3-1.5), the fuel partial pressure $p_{0,f}$ varies from 0.02 to 0.095 atm, whilst the resultant $T_{sat}$ (hence $T_{g,0}$) ranges from 277.8 to 304.6 K. Such small temperature differences are expected to have negligible effects on kernel development.

In this work, the initial droplet diameter, $d_0$, ranges from 5 to 15 μm. Before ignition, the fuel droplets are static ($u_{d,0} = 0$) and uniformly distributed in the domain. The initial droplet temperature, $T_{d,0}$, is the same as the gas temperature. Moreover, the initial number density of the fuel droplets, $N_{d,0}$, can be estimated from [50]

$$N_{d0} = \frac{\phi_l}{\phi_g + AF_s} \frac{p_{g,0} M_d}{RT_{g,0}} / (\pi \rho_d d_0^3 / 6). \qquad (26)$$

Zero-gradient conditions for species mass fractions, temperature, and liquid phase variables (diameter, temperature, and number density) are enforced at both spherical center ($r = 0$) and right boundary ($r = r_w$). Zero velocities of the gas and droplets are applied for two boundaries. To mimic the energy deposition from a spark, a source term is added for the gas phase energy equation (i.e., $S_I$ in Eq. 1) [51]

$$q_{ig}(r,t) = \begin{cases} \frac{E_{ig}}{\pi^{1.5} r_{ig}^3 \tau_{ig}} \exp\left[-\left(\frac{r}{r_{ig}}\right)^2\right], & if \ t < \tau_{ig} \\ 0, & if \ t \geq \tau_{ig} \end{cases} \qquad (27)$$

where $E_{ig}$ is the ignition energy, whilst $r_{ig}$ and $\tau_{ig}$ are the spark size and duration, respectively. In this study, $\tau_{ig} = 400\ \mu s$ and $r_{ig} = 400\ \mu m$ are assumed [52,53].

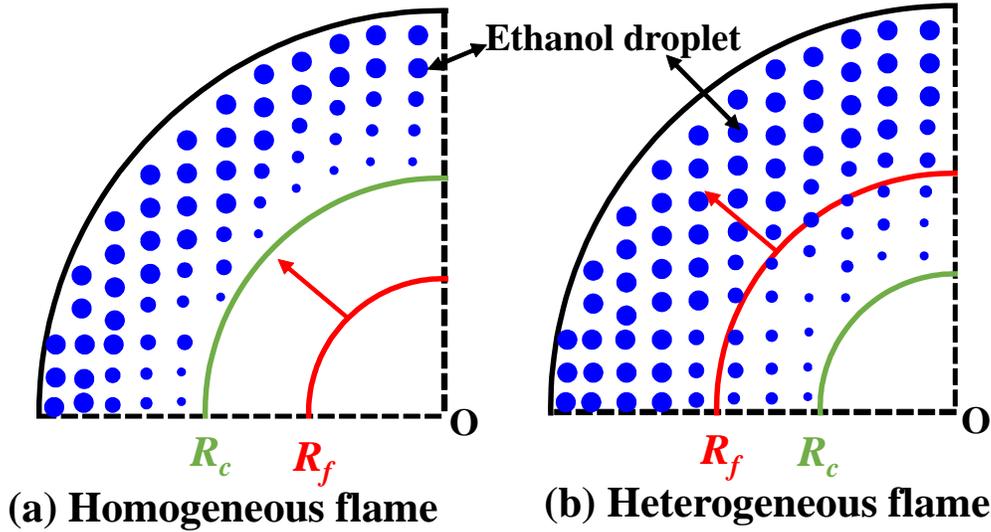

Fig. 1. Schematic of (a) homogeneous and (b) heterogeneous spray flames. Red arrow indicates flame propagates direction. Circles: fuel droplets.

According to the previous studies [13,14,19,54], interactions between evaporating droplets and reaction front significantly modulate the flame dynamics. In our recent theoretical work [13,14] and numerical studies [10] on laminar spray flames, we introduce a concept of evaporation completion front (ECF), $R_c$, where the droplets are just completely vaporized (see Fig. 1, identified as the location where the droplet diameter is critically less than a threshold value, e.g., 0.1 μm). The boundary at which the droplets are fully evaporated is observed through laser sheet imaging of spray flames, e.g., in Refs. [15,16]. Besides, the flame front (FF), $R_f$, is defined as the location with maximum heat release rate (HRR). If the front distance $\Delta R = R_c - R_f > 0$, then the ECF lies at the unburned zone, indicating that the mixture at the FF is gaseous, i.e., fuel vapor and air (droplets already fully gasified). This is termed as a *homogeneous* (HM) spray flame (Fig. 1a), correspond to Scenario I discussed in Section 1. If $\Delta R < 0$, the ECF is behind the FF. Hence, the mixture at the FF is composed of evaporating droplets, vapor, and air. This is deemed a *heterogeneous* (HT) spray flame (Fig. 1b), i.e., Scenario II.

To quantify the evolving gas composition due to evaporating sprays in both HM and HT cases, an element-based effective ER [55] is calculated for the gas phase, i.e.,

$$\phi_e = [3(\tfrac{X_C}{2} + \tfrac{X_H}{6})]/[X_O - \tfrac{1}{2}(\tfrac{X_C}{2} + \tfrac{X_H}{6})] \tag{28}$$

where $X$ denotes the mole fraction of an element (i.e., C, H or O). We always quantify the $\phi_e$ at the FF, $\phi_{e,f}$, and in our following discussion we simply term it as flame ER.

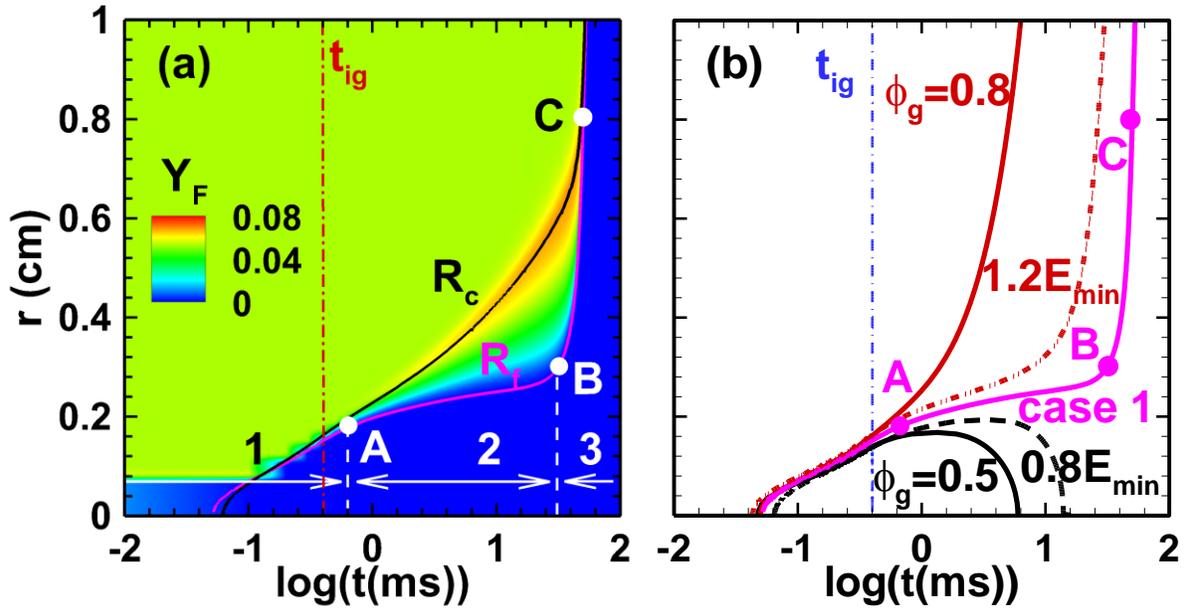

Fig. 2. (a) $r - t$ diagram of fuel mass fraction with $\phi_g = 0.5$, $\phi_l = 0.3$, and $d_0 = 5$ $\mu$m (case 1, spray flame). $E_{ig} = E_{min} = 3.3$ mJ. (b) Flame trajectories of case 1 with different ignition energies. Gaseous flames with ER = 0.5 and 0.8 are added for comparison.

## 4   Results and discussion

### 4.1   Flame kernel development in fuel-lean two-phase mixtures

The flame kernel development during spark ignition in an overall fuel-lean ethanol droplet/vapor/air mixture will be first investigated. The initial conditions are $\phi_g$ = 0.5, $\phi_l$ = 0.3, and $d_0$ = 5 $\mu$m, which is called as case 1 hereafter. Here the ignition energy is equal to the MIE, i.e., $E_{ig} = E_{min}$ = 3.3 mJ. Figure 2(a) shows the $r - t$ diagram of fuel mass fraction $Y_F$ in case 1. The corresponding evolutions of flame propagation speed ($S_b \equiv dR_f/dt$ and front distance $\Delta R$) are presented in Fig. 3(a), whilst the flame ER is shown in Fig. 3(b). For comparison, the flame trajectories and flame ER of two gaseous ethanol flames with the ER being 0.5 and 0.8 (respectively corresponding to $\phi_g$ and $\phi_{ov}$ in the spray case) are also shown in Fig. 2(b) and Fig. 3(b), respectively, and their ignition energies equal the MIE of case 1.

The kernel development in case 1 can be divided into three stages based on the evolutions of flame front and evaporation completion front, demarcated with the points A and B in Fig. 2(a). Stage 1 starts from the initial state ($t$ = 0) to the instant when the FF and ECF are decoupled, i.e., point A (0.18 cm, 0.66 ms). In this stage, both chemical reactions and droplet evaporation are largely driven by energy deposition from the spark. Hence, at the beginning of stage 1, the FF and ECF synchronously move outwardly with a small front distance $\Delta R$ (see Fig. 3a). Furthermore, the flame transits from heterogeneous ($\Delta R < 0$) to homogeneous ($\Delta R > 0$) condition. Initially, the flame ER is close to the overall ER of 0.8 (see Fig. 3b). Then, the flame ER gradually decreases in stage 1 due to the kernel decaying after the spark is terminated at $t_{ig}$.

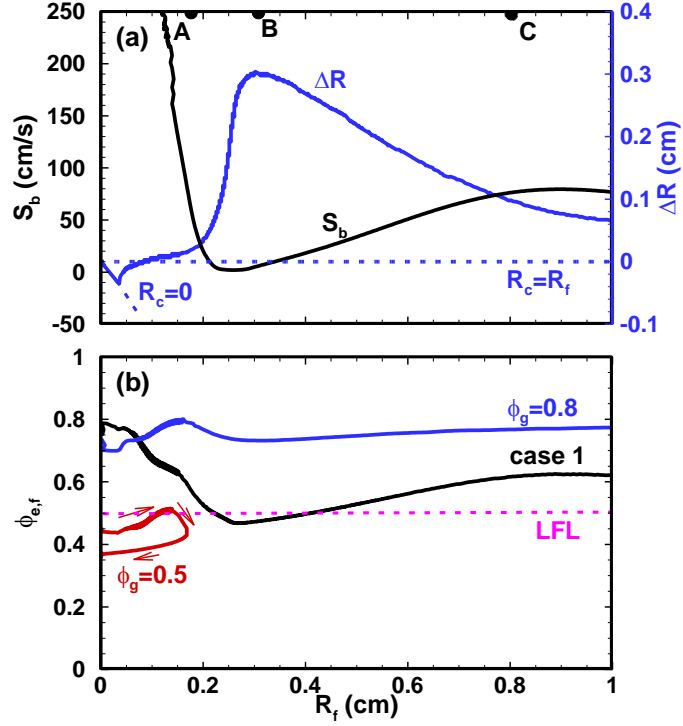

Fig. 3. (a) Change of flame propagation speed, front distance with flame radius in case 1. (b) Change of flame ER with flame radius in case 1 and two gaseous flames with ER = 0.5 and 0.8. LFL: lower flammability limit.

In stage 2 (A→ B in Fig. 2a), a quasi-stationary spherical flame (QSSF) is observed, with a gradually decreasing $S_b$. The minimum $S_b$ is 4 cm/s at $R_f = 0.27$ cm. Nonetheless, the ECF still propagates outwardly due to sufficient heating from the FF. Therefore, the front distance $\Delta R$ increases quickly. $\Delta R$ reaches the maximum at $R_f = 0.3$ cm (point B). This indicates the diffusion length for the ethanol vapor to be transported from the evaporation zone to FF is increasing and reaches the maximum value at point B. Decreased flame ER between A and B is observed (from 0.6 to 0.47), which is even smaller than the lower flammability limit (LFL) of ethanol (about 0.5 [56]), as marked in Fig. 3(b).

Beyond point B, we can observe from Fig. 3 that, $S_b$ gradually increases, while $\Delta R$ decreases. Therefore, the fuel diffusion from the ECF to FF and heat diffusion from FF to ECF are accelerated, leading to a gradually increased flame ER. Consequently, the QSSF kernel is intensified through the continuous fuel vapor supply during the QSSF period. The kernel growth is therefore accelerated, and the period after B corresponds to stage 3.

For the two gaseous flames in Fig. 2(b), it is seen that their trajectories are almost identical to case 1 in the sparking period ($t < t_{ig}$). Afterwards, the gaseous flame with $\phi_g = 0.8$ (0.5) accelerates (decelerates), leading to successful (failed) ignition event. Compared to the $\phi_g = 0.8$ gaseous flame, existence of the fuel droplets in case 1 corresponds to lower ER in the gas phase, see Fig. 3(b). This essentially induces the peculiar QSSF phenomenon for case 1. On the contrary, the ignitibility of the leaner background premixture (i.e., ER = 0.5) is improved due to the addition of fuel droplets compared to the $\phi_g = 0.5$ gaseous flame, which is demonstrated by higher flame ER in case 1 as observed from Fig. 3(b).

The ignition energy would typically affect the flame kernel trajectory in a spark ignition process [35,53]. The evolutions of the front distance and flame ER in case 1 with different ignition energies are presented in Figs. 4(a) and 4(b), respectively. As already seen from Fig. 2(b), the flame with $1.2E_{min}$ propagates faster than the MIE condition. The QSSF phenomenon is not as obvious as in case 1 with increased ignition energy. Nonetheless, for the $0.8E_{min}$ flame, ignition failure occurs (see Fig. 2c).

In Fig. 4(a), one can see that at the very early sparking period, the droplets are fully dispersed in the post-flame zone in all ignition events, which is indicated by the dashed line $R_c = 0$. After a finitely long duration, the ECF starts to deviate from the spherical center (i.e., the droplets near the spark have fully vaporized due to their longer exposure to the local hot gas). Subsequently, the kernel transits from the HT to the HM flames (crossing the $R_c = R_f$ line, see the inset of Fig. 4a). Furthermore, if $E_{ig} < E_{min}$, i.e., lines #1, failed ignition occurs. This leads to the inward movement or shrinkage of the kernel in Fig. 4(a). Meanwhile, the hot kernel continuously vaporizes fuel droplets in the pre-flame zone, resulting in increased front distance. As for the successful ignition events (lines #3 and #5), the front distance decreases with flame radius after reaching the maximum value because the flame is intensified after the QSSF period. Meanwhile, the maximum front distance decreases with ignition energy. Finally, the front distance reaches a uniform value when the flame further expands, i.e., 0.09 cm at $R_f = 2$ cm. The evolutions of the front distance in Fig. 4(a) for both failed and successful ignition events show qualitative accordance with our previous theoretical analysis on fuel spray ignition [57].

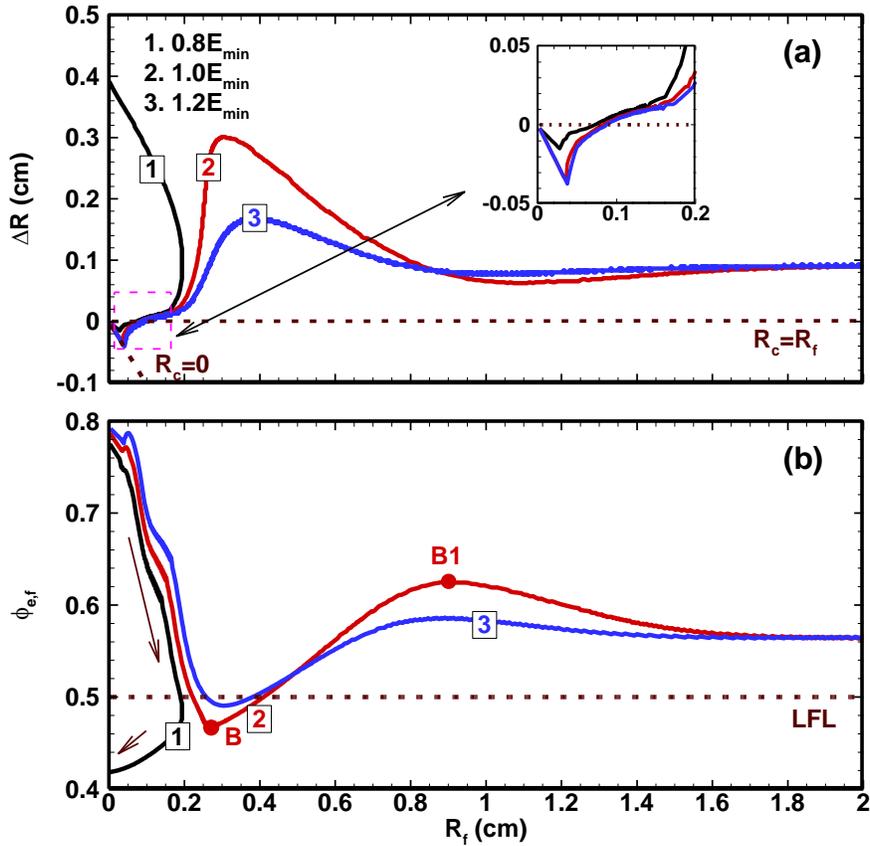

Fig. 4. Change of (a) front distance and (b) flame ER in case 1 with different ignition energies. The arrow in (b) indicates kernel propagation direction.

In Fig. 4(b), increased ignition energy leads to a slightly richer flame ER at the early stage, i.e., $R_f < 0.2$ cm. At this stage, the flame ER of each event decreases as the kernels expand. Beyond this, the flame ER of line #1 further decreases due to increased front distance (increased diffusion length). As for lines #2 and #3, the minimum flame ER (e.g., point B for line #2) increases with increased ignition energy. This is caused by the intensified evaporation process for increased ignition energy. Afterwards, the flame ER of line #3 peaks (0.63) at point B1 ($R_f = 0.85$ cm). The peak flame ER decreases with ignition energy. This is because the overdriven effect from the fuel vapor (pre-flame zone) generated during the QSSF stage is weakened with increased ignition energy (the QSSF becomes less observable with increased $E_{ig}$). Finally, the flame ER of lines #2 and #3 approaches respective constant values of 0.56 at $R_f = 2.0$ cm.

## 4.2  Flame kernel development in fuel-rich two-phase mixtures

The flame kernel development in an overall fuel-rich (i.e., $\phi_{ov} = \phi_g + \phi_l > 1.0$) ethanol droplet/vapor/air mixture will be investigated in this section. We consider a fuel-lean background gas, i.e., $\phi_g < 1.0$, and fuel-rich one ($\phi_g > 1.0$) in Sections 4.2.1 and 4.2.2, respectively.

### 4.2.1  Fuel-lean background gas ($\phi_g < 1.0$)

The selected conditions are $\phi_g = 0.3$, $\phi_l = 1.7$, and $d_0 = 15$ $\mu$m, which is called as case 2. Here the ignition energy is its MIE, i.e., $E_{ig} = E_{min} = 1.76$ mJ. Figures 5(a) and 5(b) show the $r - t$ diagrams of fuel mass fraction ($Y_F$) and HRR, respectively. Moreover, the fuel / oxidizer mass fractions and effective ER at some milestones, marked by D$_1$ - F in Fig. 5(a), are shown in Fig. 6.

In Fig. 5(a), four stages can be identified in the flame kernel development. Stage I (from $t = 0$ to point D, corresponds to the spark duration $t_{ig}$) is termed as spark ignition stage. In this stage, the *spark-ignited flame kernel* continuously expands outwardly, and the largest flame radius is observed when the spark is turned off (point D). Due to the spark effects, the HRR at the kernel is high before 0.2 ms, see point $D_1$ in Fig. 5(b). Afterwards, the HRR pronouncedly decreases, due to considerable evaporative heat loss and increased flame ER. This may result from high droplet loading, confirmed by the local high droplet volume fractions, still around 70% of the initial value, when the spark-ignited kernel reaches $D_1$ (see the detailed profiles in Fig. S1b of the supplemental material). After $D_1$, the fuel mass fraction $Y_F$ in the pre-flame zone ($r > R_f$) gradually increases to a considerable value ($> 0.08$) and forms a region rich of fuel vapor, which is seen in Fig. 5(a). This is due to decreased consumption rate of fuel vapor with weakened chemical reactions. The foregoing phenomenon is associated with the full dispersion of evaporating ethanol droplets in the entire burned area (i.e., $R_c = 0$) until point $D_2$ ($t = 0.35$ ms). At $D_2$, the droplets around the spark are fully vaporized, and an ECF (black line in Fig. 5) appears and moves outwardly but still behind the flame front, indicating an instantaneous HT flame. Due to the weakened chemical reactions from point $D_1$ to D, the maximum $Y_F$ increases from 0.03 (Fig. 6a) to 0.0825 (Fig. 6b). This is consistent with the $Y_F$ contours shown in Fig. 5(a). Moreover, the effective ER in the post-flame zone also increases from $D_1$ to D.

After the spark is terminated at D, the kernel shrinks towards E, and this is stage II. In this stage, decaying of the spark-ignited kernel is caused by the strong evaporative heat loss. Meanwhile, the flame kernel transits from HT to HM state (i.e., the mixture composition at the FF changes) after point $D_3$ ($t = 0.55$ ms). Moreover, the ECF decelerates, which is particularly evident as the ECF is not moving around $r = 0.1$ cm in Fig. 5**Error! Reference source not found.** after $D_3$. This is because the droplet evaporation slows down, due to weakened kernel and reduced heat diffusion from the flame with increased front distance. Nonetheless, the droplets continue vaporizing from D to E, leading to a continuous accumulation of the fuel vapor, transported to the decaying kernel (hence make the latter survive). This is evident from the elevated fuel vapor mass fraction $Y_F$ from point D (Fig. 6b) to E (Fig. 6c) near the dying kernel, e.g., $r < 0.08$ cm. Meanwhile, the oxidizer also diffuses inwardly, when comparing the $Y_O$ distributions in Figs. 6(b) and 6(c) near the kernel. This also leads to decrease of the effective ER (D → E) e.g., $r < 0.08$ cm, due to dilution of fuel vapor from increased $Y_O$.

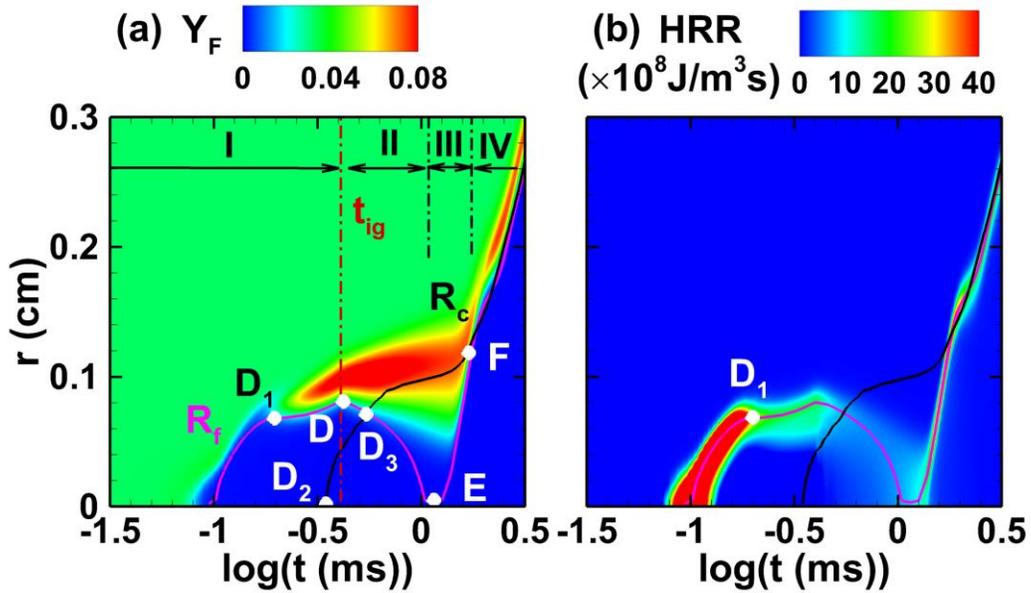

Fig. 5. $r - t$ diagram of (a) fuel vapor mass fraction and (b) heat release rate with $\phi_g = 0.3$, $\phi_l = 1.7$, $d_0 = 15$ μm. $E_{ig} = 1.76$ mJ. Solid line: trajectories of the flame front (pink) or evaporation completion front (black).

The kernel is initiated again at E, and beyond that, a new *re-ignited kernel*, different from the previous spark-ignited one, expands outwardly again. This is the re-ignition stage (from E to F, stage III). In this stage, the kernel expands outwardly at a faster speed than the ECF. Moreover, the chemical reactions at the re-ignited kernel gradually get strengthened. At point F, the flame ER is the smallest compared to the effective ER at other locations in Fig. 6(d). Since F, the ECF and re-ignited kernel couple again with a smaller front distance compared to the former stages. This stage corresponds to stage IV, i.e., flame propagation in fuel sprays.

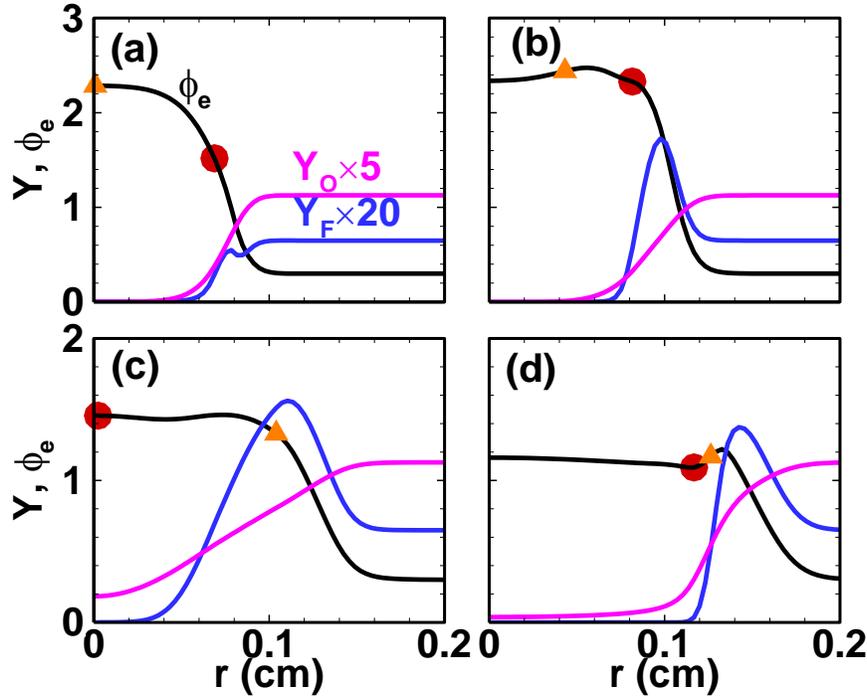

Fig. 6. Spatial distributions of fuel / oxidizer mass fractions and effective equivalence ratio at (a) $D_1$, (b) D, (c) E, and (d) F. Triangle: evaporation completion front. Circle: flame front.

To further interpret the effects of fuel sprays on the re-ignition transient, the flame trajectories of case 2 and two gaseous flames (ER = 0.3 and 2.0, respectively correspond to $\phi_g$ and $\phi_{ov}$ of case 2) are compared in Fig. 7(a). The respective flame ERs are presented in Fig. 7(b). The ignition energies of the two gaseous flames are the MIE of case 2. In Fig. 7(a), failed and successful ignition events are respectively observed for gaseous flames with the ER of 0.3 and 2.0. It is not surprising to see ignition failure when the ER is 0.3, because the flame ERs are well below the LFLF. Moreover, addition of fuel droplets is beneficial in enriching the lean background gas, which can be found in case 2 in Fig. 7(b). Compared to gaseous flame with $\phi_g = 2.0$, the flame ER of case 2 is even higher around point D, which is about 2.5, close to the upper flammability limit of ethanol (around 2.9 [56]). This is because the fuel vapor from droplet evaporation considerably affects the flame ER for the HT spray flame. The higher flame ER around D leads to lower ignitability compared to gaseous flame with $\phi_g = 2.0$.

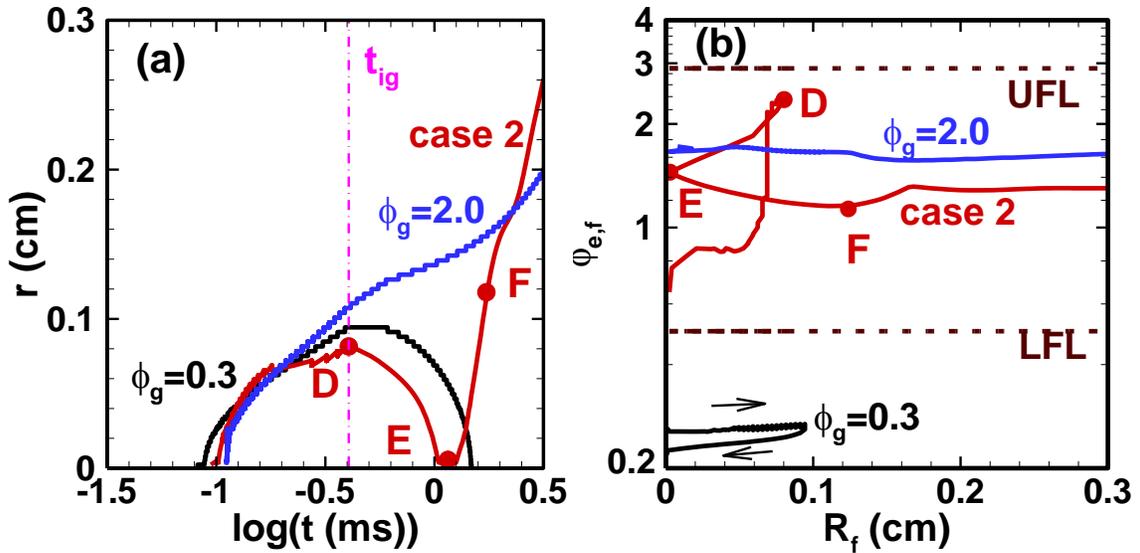

Fig. 7. (a) Flame trajectories and (b) flame ER of case 2 and two gaseous flames with $\phi_g = 0.3$ and 2.0. $E_{ig} = 1.76$ mJ. UFL: upper flammability limit.

### 4.2.2 Fuel-rich background gas ($\phi_g > 1.0$)

The spray flame kernel development in fuel-rich background gas will be discussed in this section and the case is parameterized by $\phi_g = 1.1$, $\phi_l = 0.9$, and $d_0 = 5~\mu m$ (termed as case 3 hereafter) when $E_{ig} = E_{min} = 0.585$ mJ. Figure 8(a) shows *r-t* diagram of fuel vapor mass fraction, and likewise the flame trajectories and flame ER of two gaseous flames (ER = 1.1 and 2.0) with the same $E_{ig}$ are shown in Figs. 8(b) and 8(c), respectively.

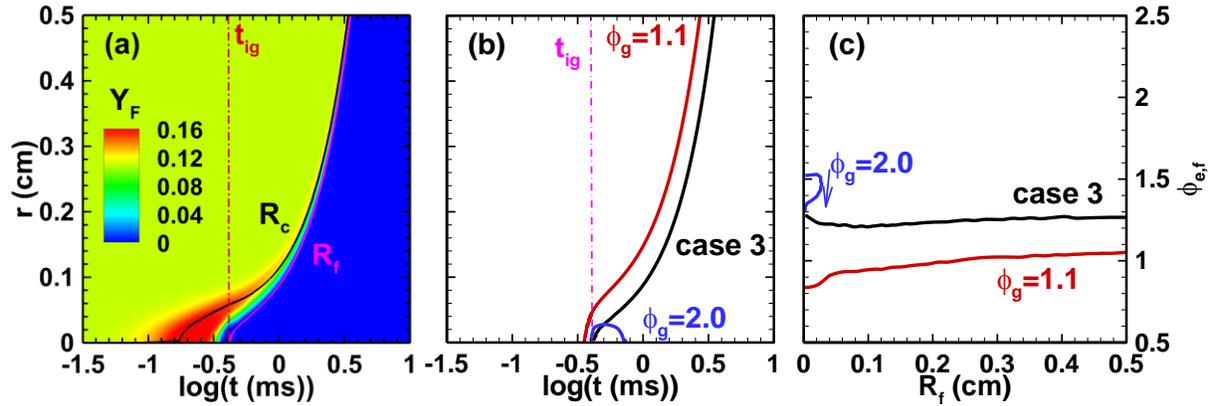

Fig. 8. (a) $r - t$ diagram of fuel vapor mass fraction with $\phi_g = 1.1$, $\phi_l = 0.9$, and $d_0 = 5$ $\mu$m (case 3). (b) Flame trajectories and (c) flame ER in case 3 and two gaseous flames with ER = 1.1 and 2.0. $E_{ig} = 0.585$ mJ.

It is seen from Fig. 8(a) that the flame kernel development differs from cases 1 and 2. Either QSSF or re-ignition phenomenon is not observed due to the near-stoichiometric gas ER (1.1) and fine droplets (5 $\mu$m). Meanwhile, in this case, the igniting kernel always grows in the HM mixture, although the front distance decreases with time. Compared to the ER = 1.1 gaseous flame, addition of fuel sprays delays the formation of igniting kernel and leads to lower flame propagation speed. This indicates that for the fuel-rich background gas, dispersed droplets near the spark deteriorate the ignitibility of the local mixture due to evaporative heat loss and/or richer flame ER from droplet evaporation [58,59], as also seen from Fig. 9(c). Moreover, the MIE of case 3 cannot ignite the gas mixture with ER = 2.0, which is evident from the flame ERs in Fig. 8(c).

Likewise, for case 4 ($\phi_g = 1.1$, $\phi_l = 0.9$, and $d_0 = 15$ $\mu$m), the $r - t$ diagram of fuel vapor mass fraction is shown in Fig. 9(a) when $E_{ig} = E_{min} = 1.16$ mJ. The comparisons with two gaseous cases are given in Figs. 9(b) and 9(c) for kernel trajectories and flame ER, respectively. From Fig. 9(a), the coarse droplets (15 μm) induce the re-ignition phenomenon, which is similar to case 2. It is seen from Figs. 9(b) and 9(c) that the difference between case 4 and the gaseous flame with ER = 1.1 is identical to that for case 3. Furthermore, a richer flame ER around the re-ignited point of case 4 is observed compared to gaseous flame with ER = 2.0. Therefore, case 4 is more difficult to be ignited than the ER = 2.0 gaseous flame, consistent with the observations from case 2.

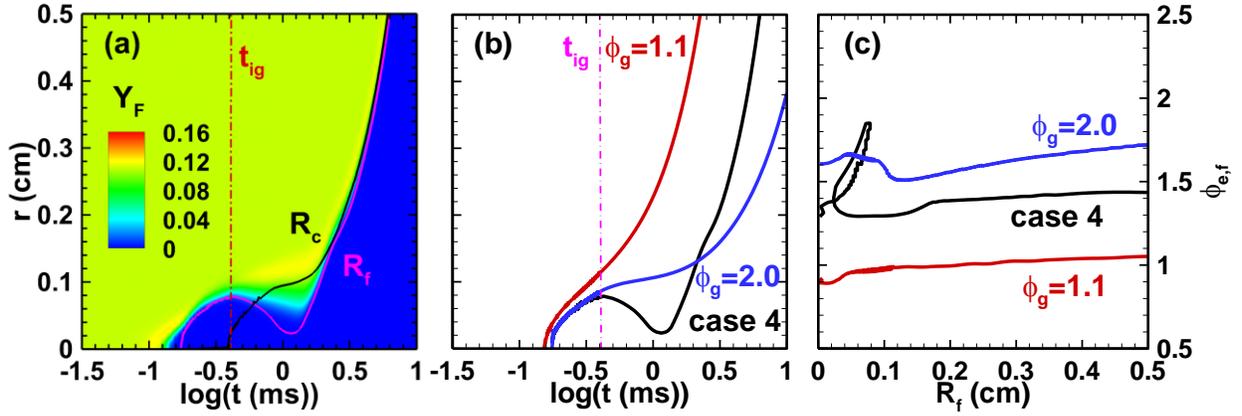

Fig. 9. (a) $r - t$ diagram of fuel vapor mass fraction with $\phi_g = 1.1$, $\phi_l = 0.9$, and $d_0 = 15$ μm (case 4). (b) Flame trajectories and (c) flame ER in case 4 and two gaseous flames with ER = 1.1 and 2.0. $E_{ig} = 1.16$ mJ.

For fuel-rich background gas in this section (cases 2 – 4), it should be noted that the ignition energy also affects the evolutions of front distance ($\Delta R$) and flame ER, particularly at the early stage of the ignition event (see details in Figs. S2 – S4 in the supplemental material). For instance, larger ignition energy $E_{ig}$ leads to a larger flame radius when the ECF starts to deviate from the spark location in cases 2 and 4. This is reasonable because the embryonic kernel expands faster with a larger $E_{ig}$. Moreover, the flame ER is also increased due to a stronger flame kernel induced by a larger $E_{ig}$ in cases 2 – 4. This is attributed to the intensified evaporation process from increased spark ignition $E_{ig}$. Meanwhile, the front distance is decreased with increased ignition energy in the kernel period. However, when the spark effects fade, the flame ER and front distance are almost not affected by the ignition energy. This is similar to the findings from Fig. 4 for spray flame in the fuel-lean background gas.

**4.3    Relations between flame kernel dynamics and minimum ignition energy**

In this section, we will investigate the minimum ignition energy of partially pre-vaporized ethanol sprays under different background gas ERs and droplet diameters. Figures 10(a) - 10(f) show the change of the MIE with overall ER. Six gas ERs are considered, i.e., $\phi_g$ = 0.3, 0.5, 0.7, 1.1, 1.5 and 2.0, respectively. Each plot shows the results with two droplet diameters, i.e., $d_0$ = 5 and 15 $\mu$m. In this study, the MIE is determined from trial-and-error simulations, with errors less than 2%.

For fuel sprays in fuel-lean background gas, i.e., $\phi_g$ = 0.3, 0.5 and 0.7 in Figs. 10(a) -10(c), the MIE firstly decreases with overall ER $\phi_{ov}$, starting from the leftmost point of the curve. For instance, in Fig. 10(a), the MIE of the $d_0$ = 5 $\mu$m sprays decreases from 5.7 mJ to 0.585 mJ when $\phi_{ov}$ increases from 1.2 to 1.7. Case 1 (marked in Fig. 10b) with the QSSF, from Section 4.1, is from this range. As mentioned above, insufficient fuel vapor at the FF is the cause of the QSSF. Increased $\phi_{ov}$ (hence $\phi_l$, due to fixed $\phi_g$) corresponds to a richer mixture at the FF. Therefore, the MIE decreases with increased $\phi_{ov}$ ($\phi_l$) in this range. This can be termed as regime A. As discussed previously in case 1, the critical flame ER is lower or close to the LFL in this regime. Furthermore, the droplet diameter and pre-vaporization degree (i.e., $\phi_g$) are found to affect the range of this regime. As seen in Fig. 10(a), this regime moves leftward (smaller $\phi_{ov}$ range) with increased droplet diameter. This can be justified by the fact that the front distance between the droplet evaporation zone and FF decreases with droplet diameter due to slower evaporation. Therefore, the local ER at the flame kernel is greater than the unity for larger droplets under the same $\phi_{ov}$ or $\phi_g$. Meanwhile, increased background gas ER also leads to a leftward shift of regime A in $\phi_{ov}$ space if one compares Figs. 10(b) and 10(c) with Fig. 10(a). This is particularly pronounced for $d_0 = 5$ $\mu$m: $\phi_{ov}$ for regime A in Fig. 10(c) corresponds to $0.7 - 1.0$, smaller than that in Fig. 10(a).

Beyond regime A, further increasing $\phi_{ov}$, e.g., 1.7-3.0 in Fig. 10(a) for $d_0 = 5$ $\mu$m, leads to weak dependence of the MIE on overall ER (around 0.58 mJ). Compared to regime A, the fuel vapor at the FF becomes relatively sufficient due to the larger $\phi_{ov}(\phi_l)$ in this range. Meanwhile, the evaporative heat loss in this range is negligible compared to the heat released from the intensified chemical reactions. This regime is featured by a plateau value of MIE for the gaseous ethanol flame when gas ER is from 0.8 to 1.8 (see Fig. S5 in the supplemental material), where the overall ER slightly affects the MIE. This range of $\phi_{ov}$ corresponds to regime B. For fixed $\phi_g$ ($d_0$), this regime lies in a smaller $\phi_{ov}$ range with increased $d_0$ ($\phi_g$), e.g., when comparing $d_0 = 5$ and 15 $\mu$m in Fig. 10(a) and $d_0 = 5$ $\mu$m in Figs. 10(a) - 10(c).

When $\phi_{ov} \geq 3.5$ and $d_0 = 5$ $\mu$m in Fig. 10(a), the slope of the MIE−$\phi_{ov}$ curve is larger compared to regime B, which can be called as regime C. This results from increased flame ER (above unity, hence richer) or evaporative heat loss. Regime C with different droplet sizes is essentially induced by different mechanisms, i.e., increased flame ER for $d_0 = 5\mu$m, whereas increased evaporative heat loss for $d_0 = 15$ $\mu$m. Moreover, the dependence of MIE on $\phi_{ov}$ is stronger than that of the 5 $\mu$m droplets due to greater evaporative heat loss near the kernel. This is evident from the larger slope of regime C for fixed $\phi_g$, i.e., $d_0 = 5$ $\mu$m and 15 $\mu$m in Fig. 10(a).

Note that case 2 in Section 4.2.1 is from this regime, in which the kernel dying/re-ignition event is observed. Similarly, distribution of the two regimes in $\phi_{ov}$ space is also affected by the droplet diameter and background gas ER, but in a more complicated way. Explanations will be given through the regime map in Fig. 11 in a wider parameter range.

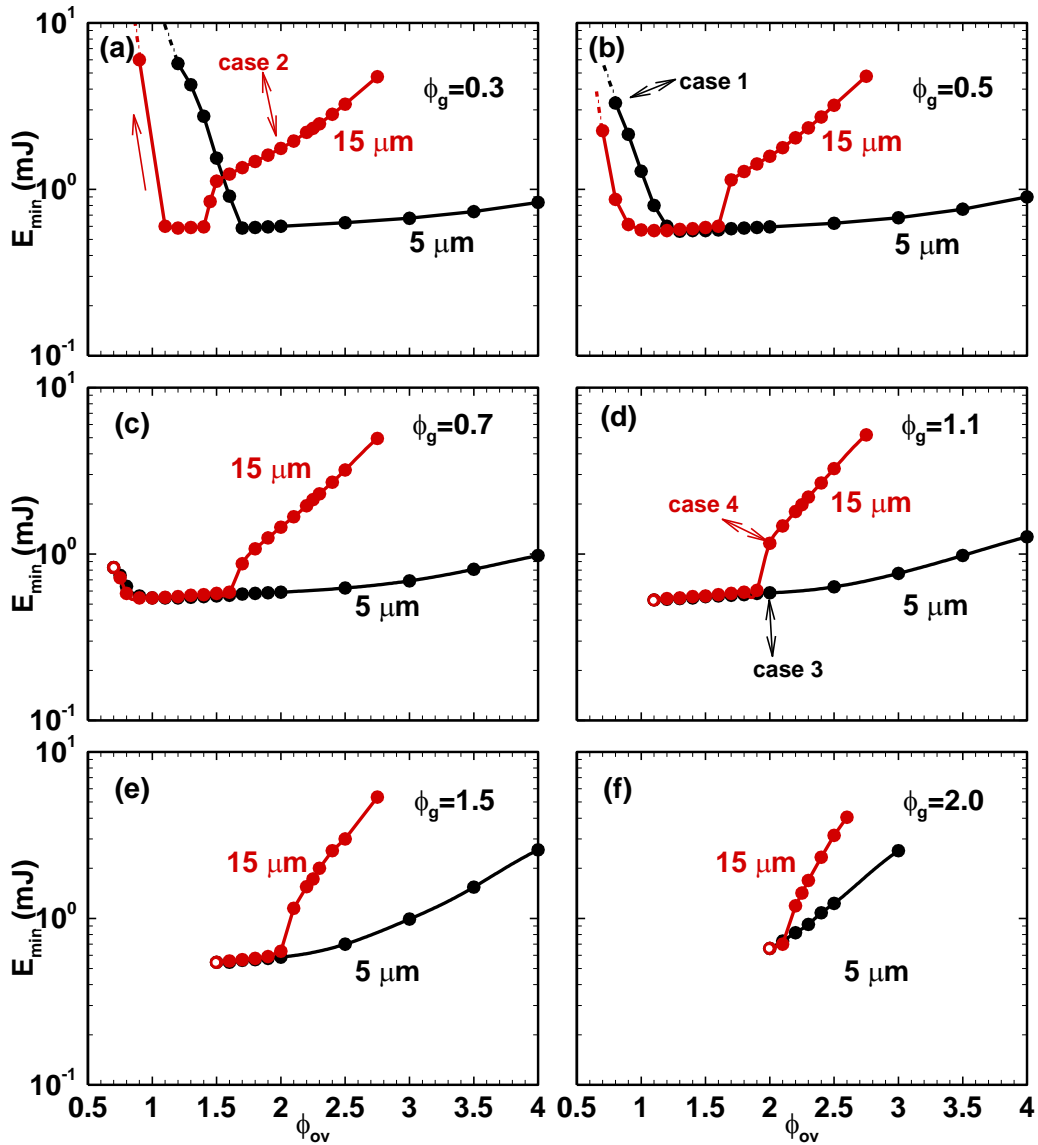

Fig. 10. Minimum ignition energy as a function of overall ER with different background gas ERs: (a) $\phi_g = 0.3$, (b) $\phi_g = 0.5$, (c) $\phi_g = 0.7$, (d) $\phi_g = 1.1$, (e) $\phi_g = 1.5$, (f) $\phi_g = 2.0$. Hollow symbol: gaseous flame under the corresponding background gas ER, i.e., $\phi_{ov} = \phi_g$.

For fuel-rich gas mixtures, e.g., $\phi_g$ = 1.1 and 1.5 in Figs. 10(d) and 10(e), the MIE monotonically increases with $\phi_{ov}$ from the gas flame point ($\phi_{ov} = \phi_g$). Regime A is not observed due to the rich background gas, because fuel vapor starvation around the igniting kernel no longer exists in these situations. Instead, only regimes B and C are shown, as illustrated in Figs. 10(c) and 10(d). In general, the MIE variations in these regimes and the underpinning mechanisms are similar to those in the fuel-lean results. Cases 3 and 4 in Section 4.2.2 can be categorized into regimes B and C, respectively (marked in Fig. 10d). It is seen that the re-ignition phenomenon also occurs in regime C (case 4) of the fuel-rich background gas. With further increased gas ER above unity, i.e., $\phi_g$ = 2.0 in Fig. 10(f), regime B also disappears or gets shortened compared to Figs. 10(d) and 10(e). This is because the background gas under $\phi_g$ = 2.0 is more difficult to be ignited than that with gas ER < 2.0 (see Fig. S5). Therefore, only regime C is presented in Fig. 10(f).

Table 1. Summary of regimes A−C of the MIE variation.

| Regime | MIE variation | Key phenomena when the ignition energy is MIE | Justification |
|---|---|---|---|
| A | Decrease with $\phi_{ov}$ | Quasi-stationary spherical flame | Flame ER near LFL |
| B | Weak dependence on $\phi_{ov}$ | Kernel formation and continuous growth | Appropriate flame ER, negligible evaporative heat loss |
| C | Increase with $\phi_{ov}$ | Re-ignition with large droplets | Flame ER above unity, considerable evaporative heat loss |

The relations between the flame kernel dynamics and MIE are further summarized in Table 1. As mentioned in Refs. [2,10,27,60,61], the dependence of the MIE on overall ER are *U*-shaped, and there exists an optimal equivalence ratio. This is consistent with the results presented in Figs. 10(a)−10(c) for partially pre-vaporized spray flame with fuel-lean background gas. Specifically, regime A(C) from Fig. 10 and Table 1 corresponds to the left (right) branch of the $U$-shaped MIE curves in Refs. [2,10,27,60,61]. Meanwhile, regime B is equivalent to the range of optimal ER in those studies. However, they do not correlate the underlying unsteady kernel dynamics to these regimes. Moreover, as for MIE shown in Figs. 10(d) – (f), limited results have been reported from previous studies, due to their relatively rich conditions of the gas phase.

To generalize the regimes of spray flame ignition, Fig. 11 summarizes three regimes of spray flame ignition in $\phi_g$-$\phi_{ov}$ space based on the MIE variation and different flame kernel dynamics. A lower ignitability limit (dash-dotted lines) is achieved, below which the mixture is not ignitable, i.e., the blue shaded zone, with the current spark radius (400 $\mu$m) and duration (400 $\mu$m) used in our simulations. Note that these spark parameters may affect this limit, but it will not be expanded for discussion in this paper. In Fig. 11, it is seen that this limit decreases with increased $\phi_g$. This is because the flame ER is increased with increased $\phi_g$ when the $\phi_{ov}$ is fixed. Moreover, increased droplet diameter from 5 to 15 $\mu$m leads to a smaller lower ignitability limit due to increased flame ER under the same gas ER. Further increasing the background gas ER, i.e., $\geq 0.7$ in Figs. 11(a) and 11(b), this ignitability limit disappears. This is reasonable because the background gas ER is generally above the LFL (around 0.5 for ethanol) in these cases. Instead, the lower boundary to achieve ignition success with a proper ignition energy is replaced by the condition of $\phi_g = \phi_{ov}$ (i.e., gaseous flame), below which is the yellow area with $\phi_g > \phi_{ov}$ (physically unachievable).

In Fig. 11(a) ($d_0 = 5$ $\mu$m), the overall ER $\phi_{ov}$ of the boundary between regime A(B) and B(C) decreases with background gas ER $\phi_g$. This is because the flame ER is the controlling parameter in the kernel dynamics and MIE variations of spray flame with small droplet diameter. Meanwhile, for fixed $\phi_{ov}$, the flame ER decreases with decreased gas ER. Similarly, when $d_0 =$ 15 μm in Fig. 11(b), the $\phi_{ov}$ of the A-B regime boundary also decreases with the gas ER. On the contrary, that of the B-C regime boundary increases with the gas ER. This is because the key factor (evaporative heat loss) is weakened with increased gas ER when $\phi_{ov}$ is fixed. Nonetheless, the B-C regime boundary in both Figs. 11(a) and 11(b) intersect with $\phi_g = \phi_{ov}$ before $\phi_g = 2.0$. This justifies why only regime C is observed at relatively large gas ER, e.g., in Fig. 10(f).

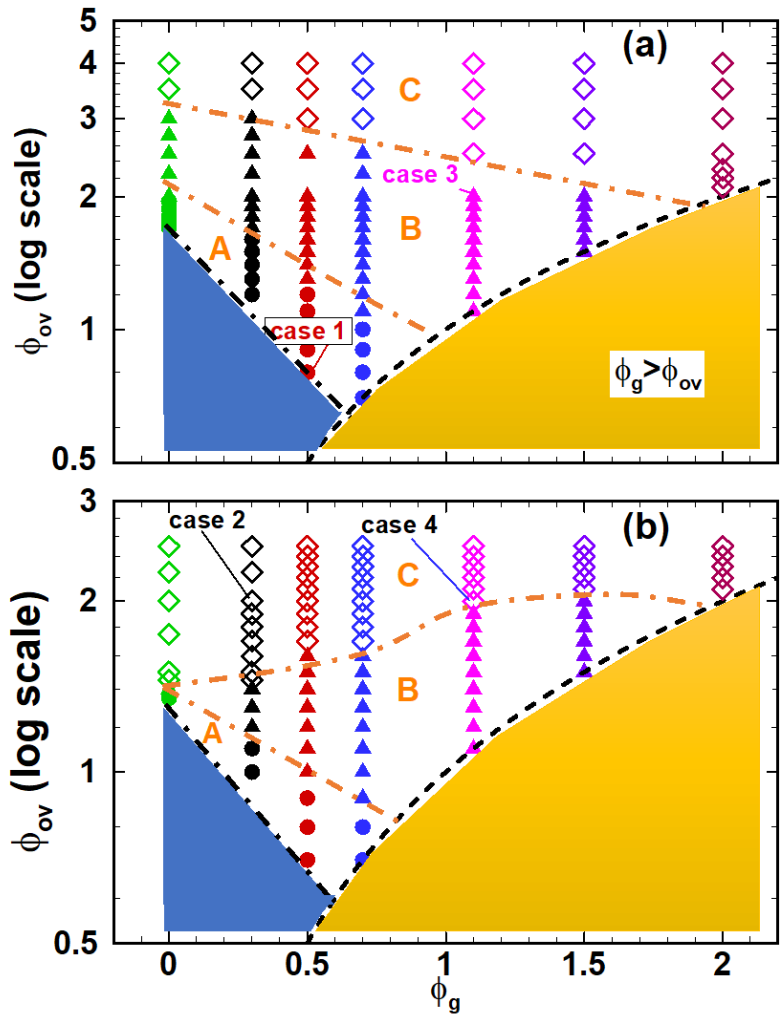

Fig. 11. Regime map of ethanol spray ignition regime in $\phi_g - \phi_{ov}$ space: (a) $d_0 = 5$ μm and (b) $d_0 = 15$ μm. Orange dash-dotted line: regime boundary. Black dashed line: $\phi_g = \phi_{ov}$. Black dash-dotted line: ignitability limit of ethanol sprays. Green symbols: pure ethanol sprays from [10].

## 5 Conclusion

This study numerically investigates spark ignition of ethanol droplet/vapor/air mixtures with an Eulerian-Eulerian method and detailed chemical mechanism. Different droplet sizes and gas/overall equivalence ratios are considered. The evaporation completion front is introduced to study the flame-droplet interaction. The gas composition at the flame front is quantified by the flame equivalence ratio.

For overall fuel-lean two-phase mixtures, the quasi-stationary spherical flame occurs due to low flame ER, which is around the lower flammability limit. The QSSF is terminated when the front distance is the largest due to the continuous fuel vapor diffusion from the evaporation zone to the flame front. Moreover, for overall fuel-rich mixtures, re-ignition of the sprays proceeds when the droplet diameter is 15 $\mu$m for fuel sprays in both lean and rich background gas. This is caused by the rich gas composition and/or considerable evaporative heat loss.

Through the comparisons with gaseous flames with gas/overall equivalence ratio, existence of fuel droplets affects flame kernel behavior and ignitability due to the heat and mass transfer between two phases. Moreover, the flame ER increases with ignition energy due to the intensified evaporation process. Meanwhile, the front distance between the flame front and evaporation completion front decreases with ignition energy. However, when the flame radius is sufficiently large, the ignition energy effects are negligible.

The minimum ignition energies of ethanol spray flames are investigated, considering different gas and overall ERs. Three regimes (A, B and C) are identified based on different dependence of the MIE on the overall ER. Different igniting kernel dynamics of these regimes are summarized. In regime A (C), the MIE decreases (increases) with increased overall ER for fixed gas ER. As for regime B, the MIE shows weak dependence on overall ER. Moreover, it is found that regime A is featured by the QSSF phenomenon, whilst regime C is correlated to the extinction/re-ignition events with large droplet size. The evolution of the three regimes with varied gas and overall ERs are compiled in parameter space of overall and gas ERs. For increased gas ER, the lower ignitibility limit decreases. With increased gas ER from fuel-lean side, regime A disappears when gas ER is close to stoichiometry. Furthermore, regime B only appears with a narrow range of overall ER when gas ER is above unity. When the background gas ER is high (i.e., higher than 2.0), only regime C exists.

**Acknowledgement**

The calculations are performed with the ASPIRE 1 Cluster from National Supercomputing Center in Singapore (https://www.nscc.sg/). QL is supported by NUS Research Scholarship. The work is supported by MOE Tier 1 grant (A-0005238-01-00).